\documentstyle[12pt]{article}

\newcommand{\be}{\begin{equation}}
\newcommand{\ee}{\end{equation}}

\newcommand{\dlt}{\delta}
\newcommand{\om}{\omega}
\newcommand{\Om}{\Omega}
\newcommand{\br}{{\bf r}}
\newcommand{\bS}{{\bf S}}
\newcommand{\bB}{{\bf B}}
\newcommand{\bt}{\beta}
\newcommand{\al}{\alpha}
\newcommand{\gm}{\gamma}
\newcommand{\Gm}{\Gamma}
\newcommand{\ra}{\rightarrow}

\newcommand{\lbd}{\lambda}
\newcommand{\prt}{\partial}

\begin{document}

\begin{center}

{\Large{\bf Absence of spin superradiance in resonatorless magnets} \\ [5mm]
V.I. Yukalov$^{1,2}$ and E.P. Yukalova$^{1,3}$} \\ [3mm]

{\it
$^1$Institut f\"ur Theoretische Physik, \\
Freie Universit\"at Berlin, Arnimallee 14, D-14195 Berlin, Germany \\ [2mm]
$^2$Bogolubov Laboratory of Theoretical Physics, \\
Joint Institute for Nuclear Research, Dubna 141980, Russia \\ [2mm]
$^3$Department of Computational Physics, Laboratory of Information
Technologies, \\
Joint Institute for Nuclear Research, Dubna 141980, Russia}

\end{center}

\vskip 2cm

{\bf Key words}: spin superradiance; collective radiation; dipole 
interactions; radiation rate; molecular magnets

\vskip 1cm

{\bf PACS}: 75.45.+j, 75.50.Xx, 75.75.+a, 75.90.+w, 76.20.+q

\newpage

\begin{abstract}

A spin system is considered with a Hamiltonian typical of molecular magnets, 
having dipole-dipole interactions and a single-site magnetic anisotropy. In 
addition, spin interactions through the common radiation field are included. 
A fully quantum-mechanical derivation of the collective radiation rate is 
presented. An effective narrowing of the dipole-dipole attenuation, due to 
high spin polarization is taken into account. The influence of the radiation 
rate on spin dynamics is carefully analysed. It is shown that this influence 
is completely negligible. No noticeable collective effects, such as 
superradiance, can appear in molecular magnets, being caused by 
electromagnetic spin radiation. Spin superradiance can arise in molecular 
magnets only when these are coupled to a resonant electric circuit, as
has been suggested earlier by one of the authors in Laser Phys. {\bf 12},
1089 (2002).

\end{abstract}

\newpage

\section{Introduction}

Superradiance is a phenomenon well known for atomic systems [1,2], where 
it occurs because of the self-organized correlation of atomic transitions 
through the common radiation field. There exists a similar phenomenon of 
spin superradiance arising in spin systems, as is reviewed in Refs. [3,4].
An accurate microscopic theory of spin superradiance has recently been
developed [5--11] demonstrating that, despite many similarities between 
the atomic and spin superradiance, these phenomena possess features that 
radically distinguish the spin from atomic superradiance [3--11]. The main 
such a fundamental difference is that the spin superradiance in real 
systems cannot be caused by their magnetodipole radiation field. The fact
that the spin radiation rate is negligible as compared to all other 
relaxation rates was, first, noticed by Bloembergen [12]. This radiation 
rate cannot yield the development of spin superradiance, as has been 
emphasized recently [13] and analysed in detail in review [4]. In order to 
realize spin superradiance, it is necessary to couple a polarized magnet 
to a resonant electric circuit. The collectivization of spin motion 
happens due to the resonator feedback field, which, thus, replaces the 
common radiation field [3--11]. The analogies and differences in atomic 
and spin superradiance have been extensively analysed in reviews [3,4,14].

Even though the physics of spin superradiance has been well established,
there have recently appeared discussions on the role of the radiation rate 
in spin relaxation. The interest to this problem has been resumed in 
connection to the studies of a novel class of materials, called molecular 
magnets, which can be highly polarized and may possess large molecular 
spins (see review [4]). The possibility of realizing spin superradiance by 
molecular magnets, coupled to a resonator, was advanced in Ref. [10], 
being based on the well developed theory [3--11]. At the same time, there 
have appeared speculations that spin superradiance in molecular magnets 
could be produced without coupling them to a resonant circuit, but in a way 
completely equivalent to atomic superradiance, so that the collectivization 
in spin motion would be due solely to their interactions through the common 
radiation field. Plausible collective effects in molecular magnets, caused 
by the radiation interaction, were discussed in Ref. [15], though the 
consideration there involved a very strong transverse magnetic field. At
this instance, it is necessary to stress that superradiance as such is 
defined as a {\it self-organized process}, without being pushed by strong 
transverse fields. This definition was given in the original paper by 
Dicke [16] and is generally accepted for both atomic [1,2,17] as well as
spin systems [3,4]. Having a generally accepted definition, there is no 
reason to change it. Therefore a process, involving strong transverse 
fields, has nothing to do with superradiance. In the best case, this could 
be {\it collective spin induction}, provided there would really be any 
collective effects. The possibility of the collective spin induction, 
caused by the molecular interactions through the electromagnetic radiation 
they are emitting, was considered in Ref. [15] in the frame of the 
phenomenological Landau-Lifshits-type equation for the classical magnetic 
moment, {\it without taking account of the dipole spin interactions}. 
However these dipole interactions do exist and are very strong in all 
really available molecular magnets.

An effort of taking into account the dipole interactions was attempted in 
Ref. [18].  Unfortunately, the authors confused the effective spin 
temperature with the real temperature of the lattice. These temperatures 
as is well known [19--21] have nothing in common, when the system is not 
in equilibrium. The lattice temperature is intrinsically positive, while 
the spin temperature is positive or negative depending upon the average 
expectation value of spin polarization. Moreover, there even exist two 
effective spin temperatures for the same spin system, the Zeeman and dipole 
effective temperatures. These temperatures are directly related to the 
average spin, which in a nonequilibrium system is a function of time. As a 
consequence, the effective spin temperatures are also functions of time and 
are defined by the Provotorov evolution equations. For inverted nonequilibrium
spins, the Zeeman temperature is negative. One should not confuse the negative 
time-dependent effective Zeeman spin temperature with the positive stationary 
real lattice temperature.

In the present paper, we give an accurate analysis of the dynamics for a 
spin system described by a microscopic Hamiltonian typical of molecular 
magnets, including the dipole interactions together with the 
electromagnetic-field interaction. All consideration is based on a fully 
quantum-mechanical picture and on well grounded mathematical methods of 
treating nonlinear spin-evolution equations. We do not couple the magnet 
to a resonant circuit, which allows us to concentrate our attention on the 
role of the radiation interactions and on the influence of the related 
radiation rate on spin motion. Our firm conclusion is that these radiation 
interactions play no role and the radiation rate is negligible, but the 
spin dynamics is governed by the dipole spin interactions. In this way, 
the spin interaction through the radiation field is absolutely unable to 
produce superradiance in molecular magnets.

\section{Electromagnetic spin interactions}

A realistic microscopic Hamiltonian for molecular magnets has the form
\be
\label{1}
\hat H= \sum_{i=1}^N \hat H_i + \frac{1}{2} \sum_{i\neq j}^N \hat H_{ij} +
\hat H_f \; .
\ee
Here the first term describes the single-spin energy, with
\be
\label{2}
\hat H_i = -\mu_0 \bS_i\cdot \bB_i - D(S_i^z)^2 \; ,
\ee
where $\mu_0=\hbar\gm_S$, $\gm_S$ being the gyromagnetic ratio for a molecule 
of spin $S$; $D$ is a single-site anisotropy parameter; the total magnetic 
field
\be
\label{3}
\bB_i = B_0{\bf e}_z + {\bf H}_i \; ,
\ee
acting on each spin, contains an external magnetic field $B_0$ and the 
radiation field ${\bf H}_i\equiv{\bf H}(\br_i,t)$. The second term in Eq. 
(1) corresponds to dipole spin interactions, with
\be
\label{4}
\hat H_{ij} = \sum_{\al\bt} D_{ij}^{\al\bt} S_i^\al S_j^\bt \; ,
\ee
where
$$
D_{ij}^{\al\bt} \equiv \frac{\mu_0^2}{r_{ij}^3}\; \left (\dlt_{\al\bt} -
3 n_{ij}^\al n_{ij}^\bt \right )
$$
is the dipolar tensor and
$$
r_{ij} \equiv |\br_{ij}|\; , \qquad {\bf n}_{ij}\equiv 
\frac{\br_{ij}}{r_{ij}} \; , \qquad {\bf r}_{ij} \equiv \br_i - \br_j \; .
$$
The last term of Eq. (1) is the Hamiltonian of electromagnetic field
\be
\label{5}
\hat H_f = \frac{1}{8\pi} \int \left ({\bf E}^2 + {\bf H}^2 \right )\;
d\br \; ,
\ee
with an electric field ${\bf E}={\bf E}(\br,t)$ and magnetic radiation 
field ${\bf H}={\bf H}(\br,t)$. The latter is expressed through the vector 
potential ${\bf A}$ as ${\bf H}=\vec\nabla\times{\bf A}$. In what follows, 
the Coulomb calibration $\vec\nabla\cdot{\bf A}=0$ is used.

To write the equations of motion for the spin operators ${\bf S}_i$ in a 
compact form, it is convenient to introduce the notation for the local 
fluctuating fields
$$
\xi_0 \equiv \frac{1}{\hbar} \; \sum_{j(\neq i)} \left ( a_{ij} S_j^z +
c_{ij}^* S_j^- + c_{ij} S_j^+ \right ) \; ,
$$
\be
\label{6}
\xi \equiv \frac{i}{\hbar} \; \sum_{j(\neq i)} \left (2 c_{ij} S_j^z -\;
\frac{1}{2}\; a_{ij} S_j^- + 2b_{ij} S_j^+ \right ) \; ,
\ee
in which $S_j^\pm$ are the ladder spin operators and
$$
a_{ij} \equiv D_{ij}^{zz} \; , \qquad b_{ij}\equiv \frac{1}{4}\left ( 
D_{ij}^{xx} - D_{ij}^{yy} - 2i D_{ij}^{xy} \right ) \; , \qquad
c_{ij} \equiv \frac{1}{2}\left ( D_{ij}^{xz} - i D_{ij}^{yz}\right ) \; .
$$
The magnetic moment operator can be represented as
\be
\label{7}
{\bf M}_i \equiv \mu_0 \bS_i = \vec\mu S_i^+ + \vec\mu^* S_i^- + 
\vec\mu_0 S_i^z \; ,
\ee
with the employed notation
$$
\vec\mu \equiv \frac{\mu_0}{2}\; \left ( {\bf e}_x - i{\bf e}_y
\right )\; , \qquad \vec\mu_0 \equiv\mu_0{\bf e}_z \; .
$$
The latter vectors enjoy the properties
$$
|\vec\mu|^2 = \frac{\mu_0^2}{2} \; , \qquad {\vec\mu^2} = 0 \; , \qquad
\vec\mu \cdot\vec\mu_0 = 0 \; .
$$
And two  more important notations are the Zeeman frequency
\be
\label{8}
\om_0 \equiv -\; \frac{1}{\hbar}\; \mu_0 B_0
\ee
and the effective field
\be
\label{9}
f \equiv -\; \frac{i}{\hbar} \; 2\vec\mu \cdot{\bf H}_i + \xi \; .
\ee
Then the equations of motion for the spin operators read as
$$
\frac{dS_i^-}{dt} = - i\left ( \om_0 -\; \frac{1}{\hbar}\; 
\vec\mu_0 \cdot {\bf H}_i + \xi_0 \right ) S_i^- + fS_i^z + 
\frac{i}{\hbar}\; D\left ( S_i^- S_i^z + S_i^z S_i^- \right ) \; , 
$$
\be
\label{10}
\frac{dS_i^z}{dt} =-\; \frac{1}{2}\left ( f^+ S_i^- + S_i^+ f\right ) \; .
\ee

The equations for the electromagnetic field are
\be
\label{11}
\frac{1}{c}\; \frac{\prt{\bf E}}{\prt t} = \vec\nabla\times{\bf H} -\;
\frac{4\pi}{c}\; {\bf j} \; , \qquad
\frac{1}{c}\; \frac{\prt{\bf A}}{\prt t} = -{\bf E} \; ,
\ee
where the current density is
\be
\label{12}
{\bf j} = - c \sum_{i=1}^N {\bf M}_i \times\vec\nabla\;\dlt(\br-\br') \; .
\ee
From Eqs. (11) one has the equation
\be
\label{13}
\left (\nabla^2 -\; \frac{1}{c^2}\; \frac{\prt^2}{\prt t^2} \right )
{\bf A} = -\; \frac{4\pi}{c}\;{\bf j} \; ,
\ee
whose solution is
\be
\label{14}
{\bf A}(\br,t) = \frac{1}{c} \; \int {\bf j}\left (\br',t-\; 
\frac{|\br-\br'|}{c} \right ) \frac{d\br'}{|\br-\br'|} \; .
\ee
The average vacuum fluctuations are assumed to be zero. The vector potential 
(14), with the current density (12), can be represented as the sum
\be
\label{15}
{\bf A}_i ={\bf A}_i^- + {\bf A}_i^+ + {\bf A}_i' \; ,
\ee
in which
$$
{\bf A}_i^- = -\sum_j \left ( 1 + \frac{1}{c}\; \frac{\prt}{\prt t}
\right ) \frac{\br_{ij}}{r_{ij}^3} \times \vec\mu^* \; S_j^- \left (
t -\; \frac{r_{ij}}{c} \right ) \; ,
$$
\be
\label{16}
{\bf A}_i' = - \sum_j \; \frac{\br_{ij}}{r_{ij}^3} \times \vec\mu_0 \; 
S_j^z \left ( t -\; \frac{r_{ij}}{c} \right ) \; ,
\ee
where the property
$$
\frac{\prt}{\prt r}\; S_i^- \left ( t  -\; \frac{r}{c} \right ) = -\;
\frac{1}{c}\; \frac{\prt}{\prt t}\; S_i^-\left ( t -\; \frac{r}{c}\right )
$$
is used. From Eqs. (15) and (16), one gets
\be
\label{17}
{\bf H}_i = \vec\nabla_i \times {\bf A}_i = {\bf H}_i^- + {\bf H}_i^+ +
{\bf H}_i' \; ,
\ee
with the expressions
$$
{\bf H}_i^- = -\sum_j \left [ 
\frac{\vec\mu^* -(\vec\mu^*\cdot{\bf n}_{ij}){\bf n}_{ij}}{c^2\; r_{ij}}\;
\frac{\prt^2}{\prt t^2} + 
\frac{\vec\mu^*-3(\vec\mu^*\cdot{\bf n}_{ij}){\bf n}_{ij}}{r_{ij}^3}\;
\left ( 1 + \frac{r_{ij}}{c}\; \frac{\prt}{\prt t} \right ) \right ] \;
S_j^-\left ( t - \; \frac{r_{ij}}{c}\right ) \; ,
$$
\be
\label{18}
{\bf H}_i' = - \sum_j \; 
\frac{\vec\mu_0 - 3(\vec\mu_0\cdot{\bf n}_{ij}){\bf n}_{ij}}{r_{ij}^3}\;
S_j^z\left ( t - \; \frac{r_{ij}}{c} \right ) \; .
\ee

Keeping in mind a macroscopic in all directions sample, we may simplify 
Eqs. (18) by averaging them over the spherical angle $\Om({\bf n}_{ij})$ 
related to the unit vector ${\bf n}_{ij}$. In doing this, we employ the 
integrals
$$
\frac{1}{4\pi} \; \int [\vec\mu - (\vec\mu\cdot{\bf n}){\bf n}]\; 
d\Om({\bf n}) = \frac{2}{3}\; \vec\mu \; ,
$$
$$
\frac{1}{4\pi} \; \int [\vec\mu - 3(\vec\mu\cdot{\bf n}){\bf n}]\; 
d\Om({\bf n}) = 0 \; ,
$$
$$
\frac{1}{4\pi} \; \int (\vec\mu\cdot{\bf n}){\bf n}\; d\Om({\bf n}) =
\frac{1}{3}\; \vec\mu \; .
$$
Then from Eqs. (18), we find
\be
\label{19}
{\bf H}_i^- = -\; \frac{2}{3}\; \vec\mu^* \; \sum_j \; 
\frac{1}{c^2\; r_{ij}} \; \frac{\prt^2}{\prt t^2} \; S_j^-\left ( t - \; 
\frac{r_{ij}}{c}\right )
\ee
and ${\bf H}_i'=0$.

Since we are looking for the possibility of arising coherent collective 
effects, we have to accept the first basic condition for the existence of
such effects. This necessary condition is the assumption that the 
transverse spin motion can be characterized by a well defined frequency 
$\om$, with the related wavelength $\lbd\equiv 2\pi/k$ and wave vector 
$k=\om/c$. Under this condition, the Born approximation is valid,
\be
\label{20}
S_j^-\left ( t -\; \frac{r}{c}\right ) = S_j^-(t)\Theta(ct-r) e^{ikr} \; ,
\ee
where $\Theta(\cdot)$ is a unit-step function, which allows us to rewrite 
Eq. (19) in the form
\be
\label{21}
{\bf H}_i^- = \frac{2}{3}\; k^3\vec\mu^* \; \sum_j \; 
\frac{\exp(ikr_{ij})}{kr_{ij}} \; \Theta(ct-r_{ij}) S_j^-(t) \; .
\ee
The magnetic field (17), with taking account of Eq. (21) and of the properties
$\vec\mu\cdot{\bf H}_i'=0$ and $\vec\mu_0\cdot{\bf H}_i^-=0$, has to be
substituted to the effective force (9) entering the spin equations of 
motion (10). The latter equations can be treated by the scale separation 
approach [5--10,14,22], whose mathematical foundation is based on the 
averaging techniques [23]. In this approach, one treats the local 
fluctuating fields (6) as random variables, which allows for the account 
of quantum spin fluctuations due to dipole interactions. Then the 
equations of motion (10) are to be averaged over the spin degrees of 
freedom. To this end, we define the averages for the {\it transverse spin}
\be
\label{22}
u(\br,t) \equiv \frac{1}{S}\; < S^-(\br,t)>\; ,
\ee
{\it coherence intensity}
\be
\label{23}
w(\br,t) \equiv \frac{1}{S^2} \; < S^+(\br,t)S^-(\br+0,t)>\; ,
\ee
and {\it spin polarization}
\be
\label{24}
s(\br,t) \equiv \frac{1}{S} \; <S^z(\br,t)> \; .
\ee

To proceed further, one needs to invoke the second fundamental condition 
for the occurrence of collective effects. This is the existence of the 
regions of strongly correlated spins, which are called {\it spin packets}. 
Really, without a substantial correlation, no collective effects can arise. 
The existence of a strongly correlated spin motion in some regions of the 
sample implies that the spin motion inside each of the regions is coherent, 
while the correlation between different {\it regions of correlation} is 
either absent or very weak. Let the characteristic length of a correlated 
region be $L_c$, volume $V_c=L_c^3$, and the number of spins inside it $N_c$. 
This characteristic length has to satisfy the inequality
\be
\label{25}
kL_c \ll 1 \; .
\ee
If inequality (25) does not hold, then all spins in sum (21) oscillate 
independently of each other, with their correlation limited by only nearest 
neighbours, which follows from the fast spatial oscillations of the kernel 
in sum (21). But under condition (25), this kernel is practically constant, 
and all spins in the region $V_c$ can be synchronized. When the wavelength 
$\lbd\gg L$ is much larger than the system length, then $L_c=L$, and the 
whole system radiates coherently. This, however, is not compulsory. The 
wavelength $\lbd$ can be much smaller than $L$. In that case, the system 
separates into several correlated regions, which radiate almost independently 
from each other. As a result, the total radiation pulse has an oscillatory 
behaviour. This is a well known picture in optical coherent radiation [1,2]. 
Lasers with a large aperture always radiate not by a unique beam but by 
a bunch of filaments [24--28]. The same is true for spin systems. When 
$\lbd\ll L$, the sample is divided into correlated regions of size $L_c\ll L$,
but such that $kL_c\ll 1$.

Inside of a correlated region, the average $<S_j^->$ weakly depends on the 
site $j$, so that this average can be taken out of the sum. Thus, we come 
to the definition of the {\it collective radiation rate}
\be
\label{26}
\gm_r \equiv \gm_0\; \sum_j^{N_c} \; \frac{\sin(kr_{ij})}{kr_{ij}}\; 
\Theta(ct-r_{ij})
\ee
and of the {\it collective frequency shift}
\be
\label{27}
\dlt\om \equiv \gm_0 \; \sum_j^{N_c} \; \frac{\cos(kr_{ij})}{kr_{ij}}\;
\Theta(ct-r_{ij}) \; ,
\ee
in which
\be
\label{28}
\gm_0 \equiv \frac{2}{3\hbar}\; \mu_0^2 Sk^3
\ee
is the single-spin natural width. Inside such a correlated region, one has
$$
-\; \frac{i}{\hbar}\; 2\; <\vec\mu\cdot{\bf H}_i > \; = 
(\gm_r -i\dlt\om) u \; .
$$

Averaging the spin equations of motion (10) over the spin degrees of freedom, 
we decouple the binary spin expressions $<S_i^\al S_j^\bt>$, for $i\neq j$, 
as $<S_i^\al><S_j^\bt>$, treating the local fluctuating fields (6) as random 
variables. At the same time, the spin products for the coinciding sites 
must be treated with caution. Averaging the last term in the first of Eq.
(10), we employ the decoupling
$$
<S_i^- S_i^z + S_i^z S_i^->\; = \left ( 2 -\; \frac{1}{S} \right )
< S_i^-><S_i^z> \; ,
$$
which accurately interpolates the spin behaviour between the two exactly 
known limits of $S=1/2$ and $S\ra\infty$ [4,10,29].

Using inequality (25), the collective radiation rate (26) and the collective 
frequency shift (27) can be simplified. The retardation effects, due to the 
unit-step function entering Eqs. (26) and (27), manifest themselves at the 
time scale $L_c/c$. Taking for $L_c$ the maximal value $L_c\sim 1/k$, the
retardation time can be approximated by $\om^{-1}$. Thus, Eqs. (26) and 
(27) can be represented as
\be
\label{29}
\gm_r \cong \gm_0 N_c\left ( 1  - e^{-\om t}\right ) \; , \qquad
\dlt\om \cong \gm_0\; \frac{3N_c}{2kL_c} \; \left ( 1 - e^{-\om t}
\right ) \; .
\ee
The time $1/\om$ is very short, usually it is the shortest among all other 
relaxation times. Therefore, for $\om t\gg 1$, and taking into account that
$N_c=\rho L_c^3$, we have for the collective radiation rate
\be
\label{30}
\gm_r \cong \gm_0 N_c = \frac{2}{3\hbar}\; \mu_0^2 S k^3 N_c
\ee
and for collective frequency shift
\be
\label{31}
\dlt\om \cong \frac{3\gm_r}{2k L_c} = \frac{1}{\hbar}\; 
\rho\mu_0^2 S (k L_c)^2 \; ,
\ee
where $\rho\equiv N/V=N_c/V_c$ is spin density. Expressions (30) and (31) 
were, first, found by Ginzburg [30], who used a classical picture, 
which also was discussed in Refs. [4,31]. Here we have presented a fully 
quantum-mechanical derivation of these expressions and, in addition, have 
obtained their generalizations (26), (27), and (29).

In studying spin dynamics, it is necessary to take account of the 
longitudinal and transverse relaxation rates. The former is due to 
spin-lattice interactions, and is denoted as $\gm_1$. The transverse 
relaxation rate is caused by the presence of dipole spin interactions 
and its standard value is
\be
\label{32}
\gm_2 = \frac{n_0}{\hbar}\; \rho\mu_0^2 \sqrt{S(S+1)} \; ,
\ee
where $n_0$ is the number of nearest neighbours. Quantity (32) is derived 
[19--21] under the assumption of a small longitudinal spin polarization. 
When the latter is large, so that the average polarization (24) substantially 
differs from zero, then the effective transverse relaxation rate narrows, 
becoming dependent on $s$. This effective narrowed rate was calculated and
described in detail by Abragam and Goldman [21], who express its value
\be
\label{33}
\gm_2(s)  =\left [ \frac{C M_2^3(s)}{M_4(s) - M_2^2(s)} \right ]^{1/2}
\ee
through the moments
$$
M_2(s) = M_2(0) (1-s^2) \; , \qquad 
M_4(s) = 2.18 M_2^2(0)(1-s^2)(1-0.42 s^2) \; .
$$
The constant $C$ in Eq. (33) depends on the line shape, being $C=\pi/2$ 
for the Gaussian and $C=\pi^2/2$ for Lorentzian lines. Substituting these
moments in Eq. (33) and keeping in mind that $s^2\leq 1$, we obtain
\be
\label{34}
\gm_2(s) = \gm_2(0) (1 - s^2) \; ,
\ee
where $\gm_2(0) =\sqrt{CM_2(0)}\equiv\gm_2$. The total transverse relaxation 
rate, including the inhomogeneous broadening $\gm_2^*$, is
\be
\label{35}
\Gm_2 = \gm_2(s) +\gm_2^* \; .
\ee

Averaging Eqs. (10) over the spin degrees of freedom, we take into account 
all described attenuations. Then, using the notation for the {\it anisotropy 
frequency}
\be
\label{36}
\om_D \equiv \frac{1}{\hbar}\; (2S -1 )D
\ee
and for the {\it effective spin frequency}
\be
\label{37}
\om_s \equiv \om_0 - \om_D s \; ,
\ee
we come to the evolution equations
$$
\frac{du}{dt} = - i(\om_s +\xi_0 -i\Gm_2) u + fs \; ,
$$
$$ 
\frac{dw}{dt} = -2\Gm_2 w + (u^* f + f^* u) s \; ,
$$
\be
\label{38}
\frac{ds}{dt} = -\; \frac{1}{2}\left ( u^* f + f^* u\right ) - 
\gm_1 ( s - \zeta) \; ,
\ee
in which the effective force is
$$
f= (\gm_r -i\dlt\om) u + \xi \; .
$$
Because of the occurrence of the random variables $\xi_0$ and $\xi$, 
corresponding to local spin fluctuations, these are stochastic differential 
equations. Averaging over the random fluctuations, treated as a Gaussian 
white noise [32], with the width $\gm_3$, we follow the scale separation 
approach [5--10,14,22]. In this way, we obtain the equations for the 
guiding centers
$$
\frac{dw}{dt} = - 2(\gm_2 +\gm_2^* -\gm_2 s^2 -\gm_r s) w + 2\gm_3 s^2 \; ,
$$
\be
\label{39}
\frac{ds}{dt} = -\gm_r w - \gm_3 s -\gm_1(s -\zeta) \; ,
\ee
containing all relaxation rates discussed above.

Comparing the relaxation rates (30) and (32), we have
\be
\label{40}
\frac{\gm_r}{\gm_2} = \frac{2}{3n_0} \sqrt{\frac{S}{S+1}} \; (k L_c)^3 \; .
\ee
According to condition (25) for the existence of correlated regions, 
$kL_c\ll 1$, and since $n_0\sim 10$, the radiation rate $\gm_r$ is always 
much
smaller than the dipole relaxation rate $\gm_2$. As is clear from Eqs. (39), 
the value $\gm_r\ll \gm_2$ plays no role in spin dynamics. Even taking the 
maximal correlation length $L_c\sim 1/k$, one has $\gm_r/\gm_2\sim 0.1$,
which does not noticeably influence the spin relaxation. Equations (39) 
contain no solutions corresponding to spin superradiance.

\section{Numerical analysis and conclusion}

We analyzed numerically various solutions to Eqs. (39). The longitudinal 
relaxation rate was assumed to be small, $\gm_1\ll\gm_2$, which is usually
the case. Other relaxation rates, expressed in units of $\gm_2$, were varied
in a wide diapason: $\gm_2^*\in[0,1]$, $\gm_3\in[0,1]$, and $\gm_r\in[0,1]$.
Different initial conditions $w_0=w(0)$ and $s_0=s(0)$ were considered,
satisfying the inequality $w_0+s_0^2\leq 1$. The solutions for $\gm_r=0$ 
and $\gm_r=0.1$ were found to be practically indistinguishable. Even 
unrealistically large $\gm_r=1$ resulted in insignificant changes, as 
compared to solutions with $\gm_r=0$. The main rates, governing the spin 
evolution, are $\gm_2$, $\gm_2^*$, and $\gm_3$. In Figs. 1-4, we show the 
solutions for $w(t)$ and $s(t)$ at different relaxation rates, demonstrating 
the role of the latter, for the same initial conditions $w_0=0$ and $s_0=1$.
These figures display the effect of {\it self-induced dynamical coherence} 
caused by the existence of dipole interactions. The main term in Eq. (39), 
producing the coherence intensity, is that containing $\gm_3$. Without this
term, no maximum in $w$ appears. The term with $\gm_2 s^2$, due to the 
narrowing of the effective transverse relaxation, which results from a high 
spin polarization, is much less important and influences the spin motion by 
about $10\% $. If at the initial moment of time, an external transverse field 
imposes a noticeable coherence intensity $w_0>0$, then we have the standard 
spin induction, as shown in Fig. 5. Both the spin induction and self-induced 
dynamical coherence occur on the time scale of the order of $T_2$. Therefore, 
though these are coherent phenomena, they have nothing to do with 
superradiance which requires the pulse time to be much shorter than $T_2$. 
Also, there is no spin reversal typical of superradiance.

Recently, there have been attempts to detect the electromagnetic radiation 
generated in the avalanches of magnetization reversal in the molecular 
magnet Mn$_{12}$-acetate, caused by the inversion of an external magnetic 
field [33]. However the most precise recent experiment [34] was unable to 
detect any significant radiation at well-defined frequencies. In any case, 
whether there is some electromagnetic radiation in the process of 
repolarizing Mn$_{12}$ or it does not exist, it has nothing to do with 
superradiance.

The accurate theory, based not on a cartoon model but on a microscopic 
consideration and taking into account the main interactions always existing 
in real molecular magnets, shows that the spin interactions through the 
common radiation field can never produce spin superradiance. This interaction 
is negligible as compared to dipole interactions and does not influence 
spin motion at all. Despite some analogy of spin systems with atomic 
ensembles displaying magnetodipole radiation [35], there are principal 
differences between them. In atomic systems, all transverse relaxation rates 
are caused by the same electromagnetic interaction, while in spin systems 
there ate several types of interactions, with the direct dipole interactions
being the strongest and the most important. So, superradiance does exist 
in atomic systems. However it is unable to arise in spin systems, if these
are not coupled to a resonator electric circuit. But if a molecular magnet 
is coupled to such a resonator, then spin superradiance can become achievable,
as is suggested in Ref. [10] and described in detail in review [4].

\vskip 5mm

{\it Acknowledgement}. Financial support from the German Research Foundation 
(grant Be 142/72-1) is appreciated. One of the authors (V.I.Y.) is grateful 
to the German Research Foundation for the Mercator Professorship.

\newpage

\begin{center}
{\large{\bf Figure Captions}}
\end{center}

\vskip 2cm

{\bf Fig. 1}. Coherence intensity  $w(t)$ (solid line) and spin polarization 
$s(t)$ (dashed line) as functions of time (measured in units of $T_2\equiv
1/\gm_2$), with the initial conditions $w_0=0$ and $s_0=1$, for the 
parameters $\gm_2^*=0$ and $\gm_3=0.1$.

\vskip 1cm

{\bf Fig. 2}. Coherence intensity  (solid line) and spin polarization 
(dashed line) versus time (in units of $T_2$), with the initial conditions 
$w_0=0$ and $s_0=1$, for $\gm_2^*=0$ and $\gm_3=1$.

\vskip 1cm

{\bf Fig. 3}. Coherence intensity  (solid line) and spin polarization 
(dashed line) versus time (in units of $T_2$), with $w_0=0$ and $s_0=1$, 
for $\gm_2^*=1$ and $\gm_3=0.1$.

\vskip 1cm

{\bf Fig. 4}. Coherence intensity  (solid line) and spin polarization 
(dashed line) versus time (in units of $T_2$), with $w_0=0$ and $s_0=1$, 
for $\gm_2^*=1$ and $\gm_3=1$.

\vskip 1cm

{\bf Fig. 5}. Spin induction. Coherence intensity  (solid line) and spin 
polarization (dashed line) versus time (in units of $T_2$), for $\gm_2^*=0$ 
and $\gm_3=1$, with the initial conditions $w_0=0.75$ and $s_0=0.5$.

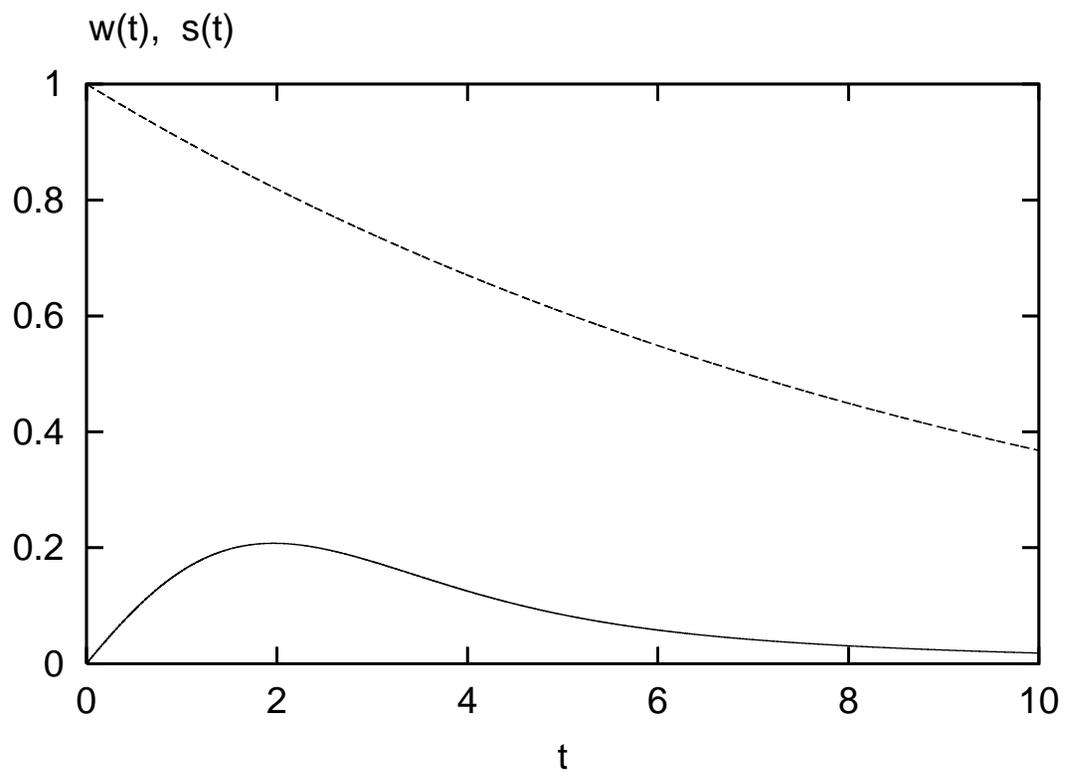
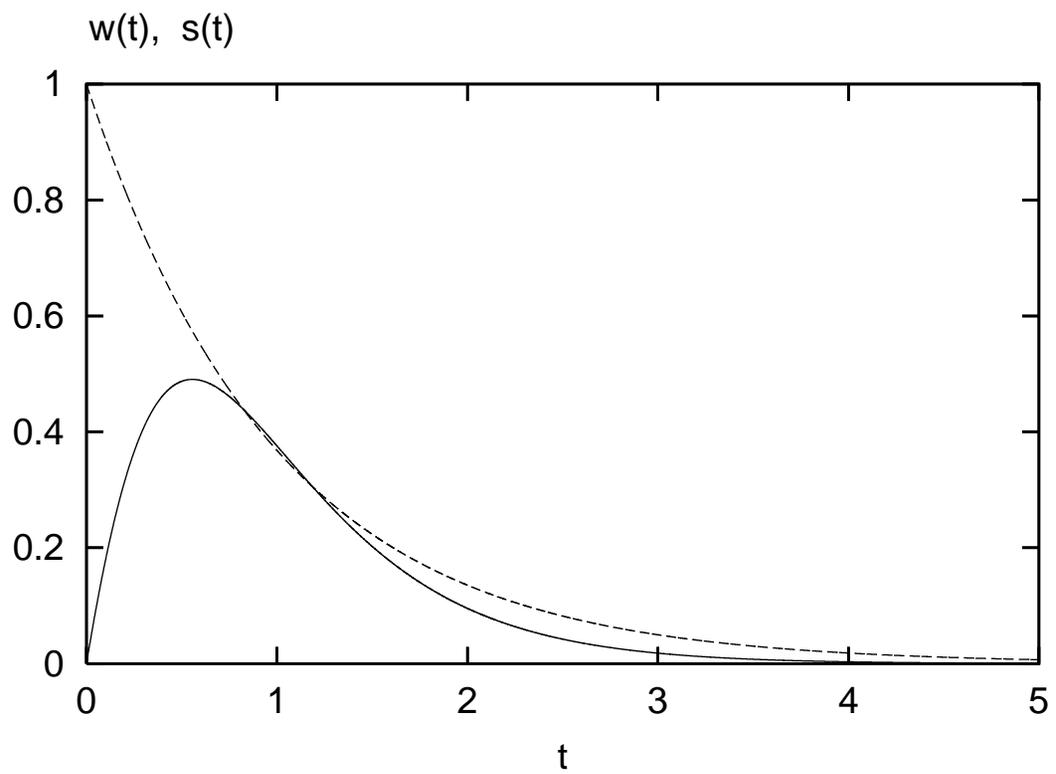
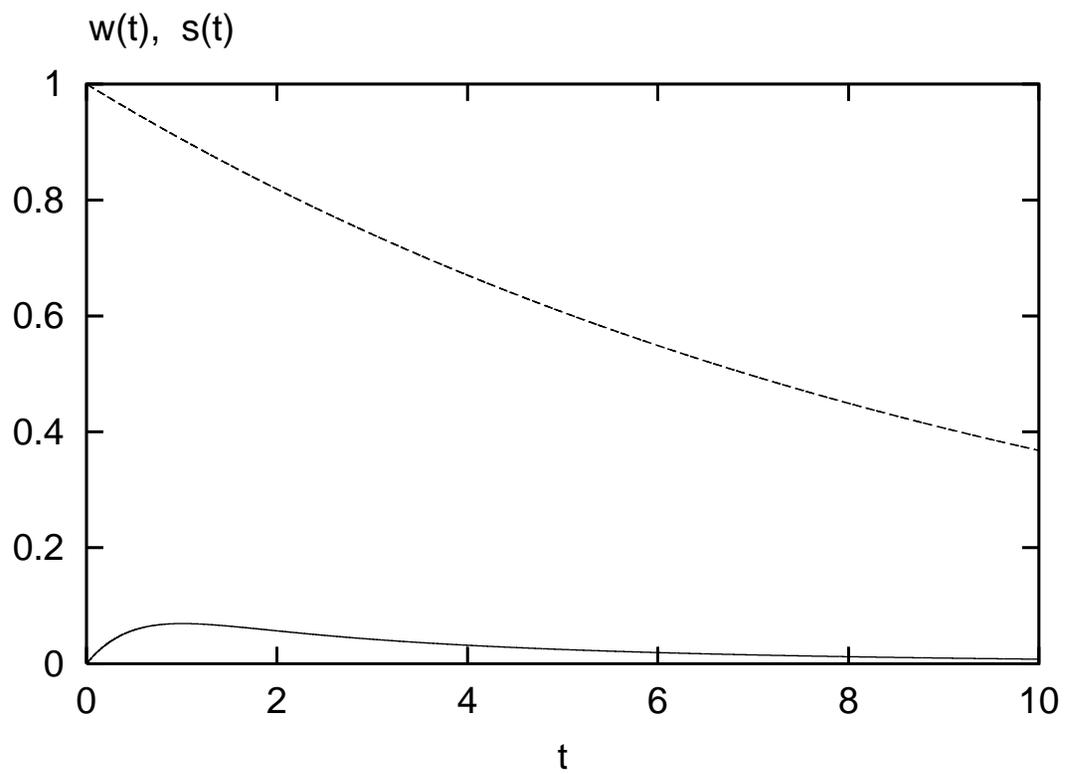
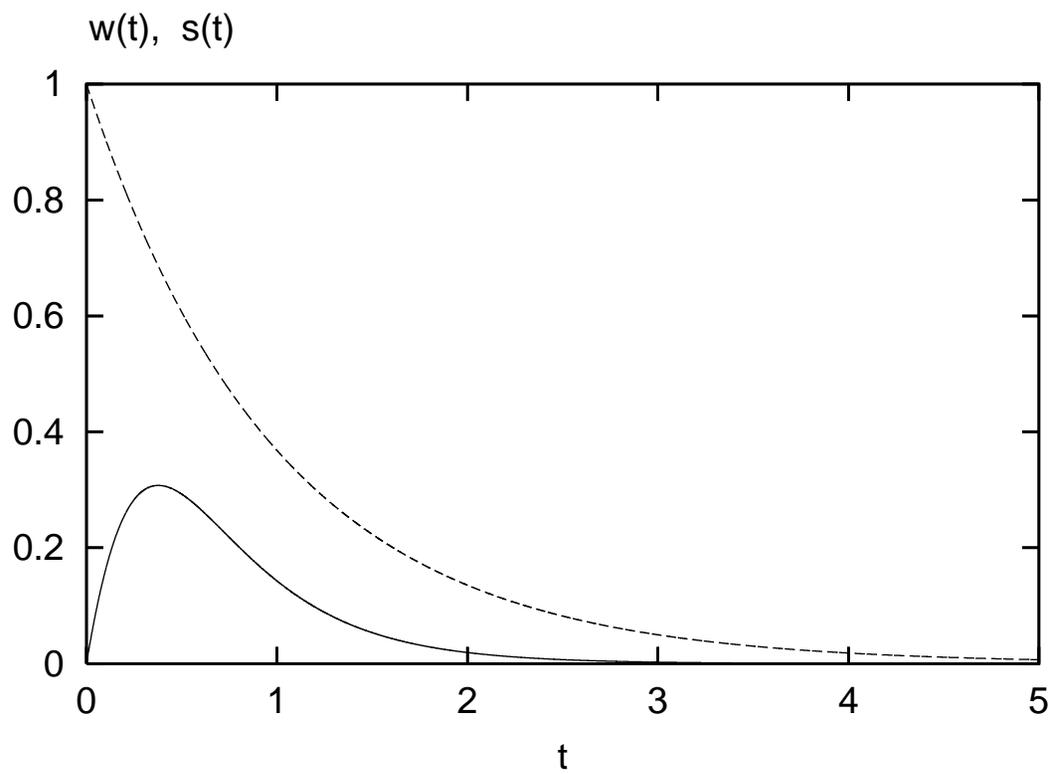
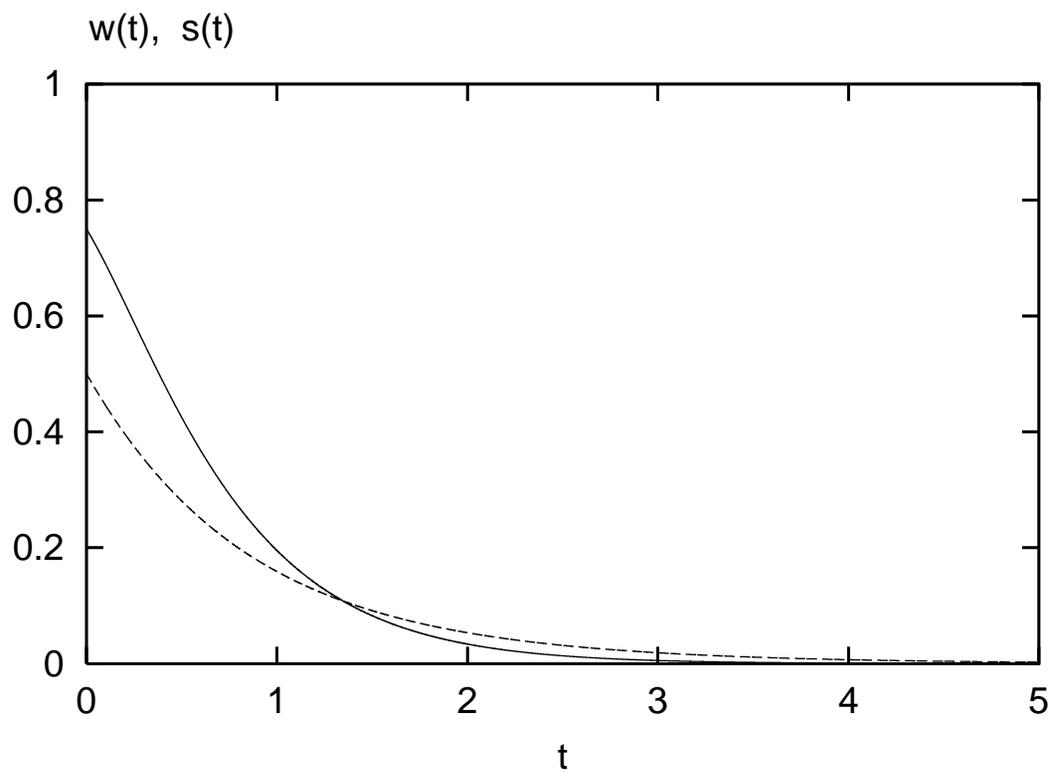
\end{document}